\documentclass[12pt,preprint]{aastex} 

\newcommand{\kms}{km~s$^{-1}$~}
\newcommand{\ie}{{\it i.e., }}
\newcommand{\etal}{{\it et al. }}
\newcommand{\eg}{{\it e.g., }}

\def\lesssim{\mathrel{\hbox{\rlap{\hbox{\lower4pt\hbox{$\sim$}}}\hbox{$<$}}}}
\def\gtrsim{\mathrel{\hbox{\rlap{\hbox{\lower4pt\hbox{$\sim$}}}\hbox{$>$}}}}

\slugcomment{To appear in Astrophysical Journal} 

\shorttitle{GRB 980425 and SN 1998bw}
\shortauthors{Weiler, \etal}

\begin{document}


\title{SN1998bw/GRB980425 and Radio Supernovae} 

\author{Kurt W. Weiler}
\affil{Naval Research Laboratory, Code 7213, Washington, DC 20375-5320} \email{weiler@rsd.nrl.navy.mil}
\author{Nino Panagia\altaffilmark{1}}
\affil{Space Telescope Science Institute, 3700 San Martin Drive,
Baltimore, MD 21218}
\email{panagia@stsci.edu}
\and
\author{Marcos J.~Montes}
\affil{NRL, Code 7213, Washington, DC 20375-5320}
\email{montes@rsd.nrl.navy.mil}

\altaffiltext{1}{Also Astrophysics Division, Space Science Department of ESA.}

\begin{abstract}
The unusual supernova SN1998bw, which is thought to be related to the $\gamma$-ray burster GRB980425, is a possible link between the two classes of objects.  Analyzing the extensive radio emission data avaliable for SN1998bw, we are able to describe its time evolution within the well established framework available for the analysis of radio emission from supernovae.  This then allows description of a number of physical properties of the object.  The radio emission can best be explained as interaction of a mildly relativistic ($\Gamma \sim1.6$) shock with a dense pre-explosion stellar wind established circumstellar medium (CSM) which is highly structured both azimuthally, in clumps or filaments, and radially, with two observed density enhancements separated by $\sim3 \times 10^{17}$\ cm.  With assumptions as to pre-explosion stellar wind conditions, it is possible to estimate that the progenitor to SN 1998bw had a mass loss rate of $\sim3.5 \times 10^{-5}\ {\rm M_\odot}~{\rm yr}^{-1}$ with at least two $\sim30\%$ increases in mass-loss rate; the most recent extending from $\sim1,600 - 4,700$\ yr before explosion and the oldest known having occurred, with possibly comparable length, $\sim12,000$\ yr before explosion.  Because of its unusual characteristics for a Type Ib/c SN, the relation of SN1998bw to GRB980425 is strengthened with consequent improvement in our understanding of these poorly understood objects.
\end{abstract}

\section{Introduction}

While generally accepted that ``most'' GRBs are extremely distant and energetic (see, e.g., \citealt{Paczynski86,Goodman86}), the discovery of GRB980425 \citep{Soffitta98} on 1998 April 25.90915 and its possible association with a bright supernova, SN1998bw at RA(J2000) = $19^h 35^m 03\fs31$, Dec(J2000) = $-52\arcdeg 50\arcmin 44\farcs7$ \citep{Tinney98}, in the relatively nearby spiral galaxy ESO~184-G82 at $z = 0.0085$ (distance $\sim40$ Mpc for $H_0 = 65$ \kms Mpc$^{-1}$) \citep{Galama98a,Galama98b,Lidman98,Tinney98,Sadler98, Galama99,Woosley99}, has introduced the possibility of a SN origin for at least some types of GRBs.  The estimated explosion date of SN1998bw in the interval 1998 April 21 -- 27 \citep{Sadler98} corresponds rather well with the time of the GRB980425 outburst.  \citet{Iwamoto98} feel that they can restrict the core collapse date for SN1998bw even more from hydrodynamical modeling of exploding C + O stars and, assuming that the SN1998bw optical light curve is energized by $^{56}$Ni decay as in Type Ia SNe, they then place the coincidence between the core collapse of SN1998bw to within +0.7/-2 days of the outburst detection of GRB980425.

Classified initially as an SN optical Type Ib \citep{Sadler98}, then Type Ic \citep{Patat98}, then peculiar Type Ic \citep{Kay98,Filippenko98}, then later, at an age of ~300 - 400 days, again as a Type Ib \citep{Patat99}, SN1998bw presents a number of optical spectral peculiarities which strengthen the suspicion that it may be the counterpart of the $\gamma$-ray burst.

When the more precise BeppoSAX NFI was pointed at the
BeppoSAX error box 10 hours after the detection of GRB980425, two X-ray sources were present \citep{Pian99}.  One of these, named S1
by \citet{Pian99}, is coincident with the position of SN 1998bw and declined slowly between 1998 April and 1998 November.  The second X-ray source, S2, which was $\sim4\arcmin$ from the position of SN 1998bw, was not (or at best only marginally with a $<3\sigma$ possible detection six days after the initial detection) detectable again in follow up observations in 1998 April, May, and November \citep{Pian99,Pian00}. 

However, although a bit of concern remains that the \citet{Pian99,Pian00} X-ray source, S2, might have been the brief afterglow from GRB980425 rather than the Pian source S1 associated with SN1998bw, \citet{Pian00} conclude that their source S1 has a "high probability" of being associated with GRB980425 and that S2 is more likely a variable field source.

In order to analyze the radio emission from SN1998bw we shall first collect, rationalize, and extend the formalism which has been so successful in understanding the radio emission from radio supernovae (RSNe) (Section 2), and apply it to SN1998bw (Section 3).  From this we are able to obtain a detailed description of the radio light curves for the object and physical parameters of the SN system (Section 4).  Finally, we compare a number of properties of Type Ib/c SNe and GRBs (Section 5).

\section{Radio Supernovae}

\subsection{Background}

A series of papers published over the past 20 years on radio supernovae (RSNe) has established the radio detection and, in a number of cases, radio evolution for approximately two dozen objects: 3 Type Ib supernovae (SNe), 3 Type Ic SNe (N.B.: Because the differences between the SN optical classes are slight -- Type Ib show strong He I absorption while Type Ic show weak He I absorption -- and there are no obvious radio differences, we shall often refer to the classes as Type Ib/c.), and the rest Type II SNe.  A much larger list of more than 100 additional SNe have low radio upper limits and can be found at: \\
{\it http://rsd-www.nrl.navy.mil/7214/weiler/kwdata/RSNtable.txt}.

In this extensive study of the radio emission from SNe, several effects have been noted: 1) No Type Ia SN has ever been detected at radio wavelengths; 2) Type Ib/c SNe are radio luminous with steep spectral indices (generally $\alpha < -1$; $S \propto \nu^{+\alpha}$) and a fast turn-on/turn-off, usually peaking at 6 cm near or before optical maximum; and 3) Type II SNe show a range of radio luminosities with flatter spectral indices (generally $\alpha > -1$) and a relatively slow turn-on/turn-off, usually peaking at 6 cm significantly after optical maximum. Type Ib/c may be fairly homogeneous in some of their radio properties while Type II, as in the optical, are quite diverse.

Measurements of the RSN multi-frequency radio light curves and their evolution with time show the density and structure of the CSM,
evidence for possible binary companions, clumpiness or filamentation in the presupernova wind, mass-loss rates and changes therein for the
presupernova stellar system and, through stellar evolution models, estimates of the ZAMS presupernova stellar mass and the stages through
which the star passed on its way to explosion.  A summary of the radio information on SNe can be found at: \\ 
{\it http://rsd-www.nrl.navy.mil/7214/weiler/sne-home.html}.

\subsection{RSN Models}

All known RSNe appear to share common properties of: 1) nonthermal synchrotron emission with high brightness temperature; 2) a decrease in absorption with time, resulting in a smooth turn-on first at shorter wavelengths and later at longer wavelengths; 3) a power-law decline of the flux density with time at each wavelength after maximum flux density (optical depth $\sim1$) is reached at that wavelength; and 4) a final, asymptotic approach of spectral index $\alpha$ ($S \propto \nu^{+\alpha}$) to an optically thin, nonthermal, constant negative value \citep{Weiler86,Weiler90}. 

\citet{Chevalier82a,Chevalier82b} has proposed that the relativistic electrons and enhanced magnetic field necessary for synchrotron emission arise from the SN blastwave interacting with a relatively high density CSM which has been ionized and heated by the initial UV/X-ray flash.  This CSM is presumed to have been established by a constant mass-loss ($\dot M$) rate, constant velocity ($w_{\rm wind}$) wind (\ie $\rho \propto \frac{\dot M}{w_{\rm wind}~r^2}$) from a massive stellar progenitor or companion.  This ionized CSM is the source of some or all of the initial absorption although \citep{Chevalier98} has proposed that synchrotron self-absorption (SSA) may play a role in some objects.  

A rapid rise in the observed radio flux density results from a decrease in these absorption processes as the radio emitting region expands and the absorption processes, either internal or along the line-of-sight, decrease.  \citet{Weiler90} have suggested that this CSM can be ``clumpy'' or ``filamentary,'' leading to a slower radio turn-on, and \citet{Montes97} have proposed at least one example for the presence of a distant ionized medium along the line-of-sight which is time independent and can cause a spectral turn-over at low radio frequencies.  In addition to clumps or filaments, the CSM may be structured with significant density irregularities such as rings, disks, shells, or gradients.

\subsection{RSN Parameterized Radio Light Curves}

Following \citet{Weiler86}, \citet{Weiler90}, and \citet{Montes97}, we adopt a parameterized model (N.B.: The notation is extended and rationalized here from previous publications.  However, the ``old'' notation of $\tau$, $\tau^{\prime}$, and $\tau^{\prime \prime}$, which has been used previously, is noted, where appropriate, for continuity.):

\begin{equation}
S(\mbox{mJy}) = K_1 \left(\frac{\nu}{\mbox{5\
GHz}}\right)^{\alpha} \left(\frac{t-t_0}{\mbox{1\ day}}\right)^{\beta}
e^{-\tau_{\rm external}} \left(\frac{1-e^{-\tau_{{\rm CSM}_{\rm clumps}}}}{\tau_{{\rm CSM}_{\rm
clumps}}}\right) \left(\frac{1-e^{-\tau_{\rm internal}}}{\tau_{\rm internal}}\right) 
\end{equation} 

\subsubsection{External Absorption:  Uniform \& Distant} 

\begin{equation}
\tau_{\rm external}  =  \tau_{{\rm CSM}_{\rm uniform}}+\tau_{\rm distant} = \tau + \tau^{\prime\prime}
\end{equation}

\begin{equation}
\tau_{{\rm CSM}_{\rm uniform}} = \tau  =  K_2
\left(\frac{\nu}{\mbox{5 GHz}}\right)^{-2.1}
\left(\frac{t-t_0}{\mbox{1\ day}}\right)^{\delta}
\end{equation}

\begin{equation}
\tau_{\rm distant}  =   \tau^{\prime\prime}  =  K_4  \left(\frac{\nu}{\mbox{5\
GHz}}\right)^{-2.1}
\end{equation} 

\subsubsection{External Absorption:  Clumpy or Filamentary} 

\begin{equation}
\tau_{{\rm CSM}_{\rm clumps}}  =   \tau^{\prime} =  K_3 \left(\frac{\nu}{\mbox{5\
GHz}}\right)^{-2.1} \left(\frac{t-t_0}{\mbox{1\
day}}\right)^{\delta^{\prime}}
\end{equation} 

\noindent with $K_1$, $K_2$, $K_3$, and $K_4$ corresponding, formally, to the flux density ($K_1$), uniform ($K_2$, $K_4$), and clumpy or filamentary ($K_3$) absorption, at 5~GHz one day after the explosion date $t_0$.  The terms $\tau_{{\rm CSM}_{\rm uniform}}$ and  $\tau_{{\rm CSM}_{\rm clumps}}$ describe the attenuation of local, uniform CSM and clumpy CSM that are near enough to the SN progenitor that they are altered by the rapidly expanding SN blastwave.  $\tau_{{\rm CSM}_{\rm uniform}}$ is produced by an ionized medium that uniformly covers the emitting source (``uniform external absorption''), and the $(1-e^{-\tau_{{\rm CSM}_{\rm clumps}}}) \tau_{{\rm CSM}_{\rm clumps}}^{-1}$ term describes the attenuation produced by an inhomogeneous medium with optical depths distributed, with equal probability, between 0 and $\tau_{{\rm CSM}_{\rm clumps}}$ (``clumpy absorption''; See \citealt{Natta84} for a more detailed discussion of attenuation in inhomogeneous media).    The $\tau_{\rm distant}$ term describes the attenuation produced by a homogeneous medium which uniformly covers the source but is so far from the SN progenitor that it is not affected by the expanding SN blastwave and is constant in time.  All absorbing media are assumed to be purely thermal, singly ionized gas which absorbs via free-free (f-f) transitions with frequency dependence $\nu^{-2.1}$ in the radio.  The parameters $\delta$ and $\delta'$ describe the time dependence of the optical depths for the local uniform and clumpy or filamentary media, respectively. 

The f-f optical depth outside the emitting region is proportional to the integral of the square of the CSM density over the radius.  Since in the simple Chevalier model the CSM density decreases as $r^{-2}$, the external optical depth will be proportional to $r^{-3}$,  and since the radius increases as a power of time, $r \propto t^m$ with $m \leq 1$ (\ie m = 1 for undecelerated blastwave expansion), it follows that the deceleration parameter, m, is

\begin{equation}
m = -\delta / 3.
\end{equation}

\noindent The model by \citet{Chevalier82a,Chevalier82b} relates $\beta$ and $\delta$ to the energy spectrum of the relativistic particles $\gamma$ ($\gamma = 2\alpha-1$) by $\delta = \alpha - \beta - 3$ so that, for cases where $K_2 = 0$ and $\delta$ is, therefore, indeterminate, we can use

\begin{equation}
m = -(\alpha - \beta - 3)/3.
\end{equation}

\subsubsection{Internal Absorption: SSA \& Mixed f-f Absorption/Nonthermal Emission} 

Since it is physically realistic and may be needed in some RSNe where radio observations have been obtained at early times and high frequencies, we have also included in Eq.~1 the possibility for an internal absorption term\footnote{Note that for simplicity we use an internal absorber attenuation of the form $\left(\frac{1-e^{-\tau_{{\rm CSM}_{\rm internal}}}}{\tau_{{\rm CSM}_{\rm internal}}}\right)$, which is appropriate for a plane-parallel geometry, instead of the more complicated expression (e.g., \citealt{Osterbrock74}) valid for the spherical case.  The assumption does not affect the quality of our analysis because, to within 5\% accuracy, the optical depth obtained with the spherical case formula is simply three-fourths of that obtained with the plane-parallel slab formula.}.  This internal absorption  ($\tau_{\rm internal}$) term may consist of two parts -- synchrotron self-absorption (SSA; $\tau_{{\rm internal}_{\rm SSA}}$), and mixed, thermal f-f absorption/non-thermal emission ($\tau_{{\rm internal}_{\rm ff}}$). 

\begin{equation}
\tau_{\rm internal}  = \tau_{\rm internal_{\rm SSA}} + \tau_{\rm internal_{\rm ff}}
\end{equation}
\begin{equation}
\tau_{\rm internal_{\rm SSA}} = K_5\left(\frac{\nu}{\mbox{5\
GHz}}\right)^{\alpha-2.5}  \left(\frac{t-t_0}{\mbox{1\
day}}\right)^{\delta^{\prime\prime}}
\end{equation}
\begin{equation}
\tau_{\rm internal_{\rm ff}}  =   K_6  \left(\frac{\nu}{\mbox{5\
GHz}}\right)^{-2.1} \left(\frac{t-t_0}{\mbox{1\
day}}\right)^{\delta^{\prime\prime\prime}}
\end{equation}

\noindent with $K_5$ and $K_6$ corresponding, formally, to the internal, non-thermal ($\nu^{\alpha - 2.5}$) SSA ($K_5$) and the internal, mixed with nonthermal emission, thermal ($\nu^{-2.1}$) f-f absorption ($K_6$), respectively, at 5~GHz one day after the explosion date $t_0$.  The parameters $\delta^{\prime \prime}$ and $\delta^{\prime \prime \prime}$ describe the time dependence of the optical depths for the SSA and f-f internal absorption components, respectively. 

A cartoon of the expected structure of an SN and its surrounding media is presented in Fig.~\ref{fig1} (see also, \citealt{Lozinskaya92}).  The radio emission is expected to arise near the blastwave \citep{Chevalier94}.

\subsection{RSN Results}

The success of the basic parameterization and model has been shown in the relatively good correspondence between the model fits and the data for all subtypes of RSNe: $\eg$ Type Ib SN1983N \citep{Weiler84}, Type Ic SN1990B \citep{VanDyk93a}, Type II SN1979C \citep{Weiler91,Weiler92a,Montes00} and SN1980K \citep{Weiler92b,Montes98}, and Type IIn SN1988Z \citep{VanDyk93,Lacey01}.  (Note that after day $\sim4000$, the evolution of the radio emission from both SN1979C and SN1980K deviates from the expected model evolution, that SN1979C shows a sinusoidal modulation in its flux density prior to day $\sim$4000, and that SN1988Z shows a double drop in its flux density, after day $\sim$ 2400 and again after day $\sim$3100.)  A more detailed discussion of radio observations of SNe and of modeling results is given in \citet{Weiler01}
 
Thus, the radio emission from SNe appears to be relatively well understood in terms of blastwave interaction with a structured CSM as described by the \cite{Chevalier82a,Chevalier82b} model and its modifications by \citet{Weiler86}, \citet{Weiler90}, and \citet{Montes97}.  For instance, the fact that the uniform external absorption exponent $\delta$ is $\sim -3$, or somewhat less, for most RSNe is evidence that the absorbing medium is generally an $\rho \propto r^{-2}$ wind as expected from a massive stellar progenitor which explodes in the red supergiant (RSG) phase.  Also, the relatively smooth behavior of the radio light curves suggests that both the external medium and the relativistic electron acceleration process are rather regular and smooth. In addition, there are a number of physical properties of SNe which we can determine from radio observations and modeling.  

\subsubsection{Mass-Loss Rate from Radio Absorption}

From the \cite{Chevalier82a,Chevalier82b} model, the turn-on of the radio emission for RSNe provides a measure of the presupernova mass-loss rate to wind velocity ratio ($\dot M/w_{\rm wind}$).  For the case in which the external absorption is entirely due to a uniform medium, using the formulation of \cite{Weiler86} Eq.~16, we can write 

\begin{eqnarray}
\frac{\dot M ({\rm M_\odot} ~ {\rm yr}^{-1})}{( w_{\rm wind} / 10\ {\rm km\ s}^{-1} )} & = &
3.0 \times 10^{-6}\ \tau_{{\rm CSM}_{{\rm uniform}}}^{0.5}\ m^{-1.5} {\left(\frac{v_{\rm i}}{10^{4}\ {\rm km\ s}^{-1}}\right)}^{1.5} \times \nonumber \\ & & {\left(\frac{t_{\rm i}}{45\ {\rm days}}\right) }^{1.5} {\left(\frac{t}{t_{\rm i}}\right) }^{1.5 m}{\left(\frac{T}{10^{4}\ {\rm K}} \right)}^{0.68}
\end{eqnarray}

\noindent Since the appearance of optical lines for measuring SN ejecta velocities is often delayed a bit relative to the time of the explosion, we arbitrarily take $t_{\rm i}$ = 45 days.  Because our observations have shown that, generally, 0.8 $\le m \le$ 1.0 and from Eq.~11 $\dot M \propto t_{\rm i}^{1.5(1-m)}$, the dependence of the calculated mass-loss rate on the date $t_{\rm i}$ of the initial ejecta velocity measurement is weak, $\dot M \propto t_{\rm i}^{<0.3}$.  Thus, we can take the best optical or VLBI velocity measurements available without worrying about the deviation of their exact measurement epoch from the assumed 45 days after explosion.  For convenience, and because many SN measurements indicate velocities of $\sim10,000$ \kms, we take $v_{\rm i} = v_{\rm blastwave} = 10,000$ \kms.  We also normally adopt values of $T = 20,000$ K, $w_{\rm wind} = 10$ \kms (which is appropriate for a RSG wind) and, generally from our best fits to the radio data for each RSN, $t = (t_{\rm 6cm\ peak} - t_0)$ days, $\tau_{{\rm CSM}_{\rm uniform}}$ from Eq. 3, and $m$ from Eq. 6 or 7, as appropriate.  

\subsubsection{Clumpiness of the Presupernova Wind}
 
In their study of the radio emission from SN1986J, \cite{Weiler90} found that the simple \cite{Chevalier82a,Chevalier82b} model could not describe the relatively slow turn-on.  They therefore added terms described mathematically by $\tau_{{\rm CSM}_{\rm clumps}}$ in Eqs.~1 and 5.  This extension greatly improved the quality of the fit and was interpreted by \cite{Weiler90} to represent the possible presence of filamentation or clumpiness in the CSM.

Such a clumpiness in the wind material was again required for modeling the radio data from SN1988Z \citep{VanDyk93} and SN1993J \citep{VanDyk94}.  Since that time, evidence for filamentation in the envelopes of SNe has also been found from optical and UV observations
(e.g., \citealt{Filippenko94,Spyromilio94}).

In the presence of a clumpy absorbing medium, the expression to derive the mass-loss rate given by Eq.~11 needs to be revised.  For the case
of $\delta=\delta'$, it is clear that the fraction of clumpy material remains constant throughout the whole wind established CSM and, therefore, the radio signal from the SN suffers an absorption $\tau_{{\rm CSM}_{\rm uniform}}$ from the uniform component of the CSM plus an additional absorption with an even probability distribution between 0 and $\tau_{{\rm CSM}_{\rm clumps}}$ from the clumpy or filamentary component of the CSM.  Since  the mass-loss rate is proportional to the square root of the optical depth, it will be enough to replace $\tau_{{\rm CSM}_{\rm uniform}}^{0.5}$ in Eq.~11 with the square root of  an effective optical depth, $\tau_{{\rm eff}}^{0.5}$ (see also \citealt{VanDyk94}), which is the appropriate average over the possible extremes of the optical depth and is defined as 

\begin{equation}
<\tau_{{\rm eff}}^{0.5}> = 0.67~ [(\tau_{{\rm CSM}_{\rm uniform}} + \tau_{{\rm CSM}_{\rm clumps}})^{1.5}-\tau_{{\rm CSM}_{\rm uniform}}^{1.5}]\tau_{{\rm CSM}_{\rm clumps}}^{-1}
\end{equation}

\noindent so that Eq.(11) can more generally be written as

\begin{eqnarray}
\frac{\dot M ({\rm M_\odot} ~ {\rm yr}^{-1})}{( w_{\rm wind} / 10\ {\rm km\ s}^{-1} )} & = &
3.0 \times 10^{-6}\ \tau_{\rm eff}^{0.5}\ m^{-1.5} {\left(\frac{v_{\rm i}}{10^{4}\ {\rm km\ s}^{-1}}\right)}^{1.5} \times \nonumber \\ & & {\left(\frac{t_{\rm i}}{45\ {\rm days}}\right) }^{1.5} {\left(\frac{t}{t_{\rm i}}\right) }^{1.5 m}{\left(\frac{T}{10^{4}\ {\rm K}} \right)}^{0.68}
\end{eqnarray}

\noindent which, in the limit of $\tau_{{\rm CSM}_{\rm clumps}} \longrightarrow 0$ approaches Eq.~11 and in the limit of $\tau_{{\rm CSM}_{\rm uniform}} \longrightarrow 0$ approaches Eq.~11 with $\tau_{{\rm CSM}_{\rm uniform}}^{0.5}$ replaced by 0.67 $\tau_{{\rm CSM}_{\rm clumps}}^{0.5}$.
 
While intermediate cases where $\delta \neq \delta^{\prime}$, $\tau_{{\rm CSM}_{\rm uniform}} \neq 0$, or $\tau_{{\rm CSM}_{\rm clumps}} \neq 0$ will yield results with larger errors, it is felt, considering other uncertainties in the assumptions, that Eqs.~11 and 13 yield reasonable estimates of the mass-loss-rates of the presupernova star.  Absorption mass-loss rate estimates from Eq. 11 or 13, as appropriate, for Type Ib/c SNe tend to be $\sim10^{-6}\ {\rm M_\odot} ~ {\rm yr}^{-1}$ while those for Type II SNe tend to be higher at $\sim10^{-4} - 10^{-5}\ {\rm M_\odot} ~ {\rm yr}^{-1}$.

\subsubsection{Mass-Loss Rate from Radio Emission}

For comparison purposes, we can also try to estimate the presupernova mass-loss rate which established the CSM by considering the radio emission directly.  From the Chevalier model and \cite{Weiler89} one can write the radio emission from an RSN in the form: 

\begin{equation}
\frac{\dot M ({\rm M_\odot} ~ {\rm yr}^{-1})}{( w_{\rm wind} / 10\ {\rm km\ s}^{-1} )} =
8.6 \times 10^{-9}\ {\left(\frac{L_{\rm 6cm~peak}}{10^{26}~{\rm ergs\ s^{-1}\ Hz^{-1}}}\right)}^{0.71} {\left(\frac{{t_{\rm 6cm\ peak}-t_0}}{({\rm days})}\right)}^{1.14}
\end{equation}

\noindent for Type Ib/c SNe and

\begin{equation}
\frac{\dot M ({\rm M_\odot} ~ {\rm yr}^{-1})}{( w_{\rm wind} / 10\ {\rm km\ s}^{-1} )} =
1.0 \times 10^{-6}\ {\left(\frac{L_{\rm 6cm\ peak}}{10^{26}~{\rm ergs\ s^{-1}\ Hz^{-1}}}\right)}^{0.54} {\left(\frac{{t_{\rm 6cm\ peak}-t_0}}{({\rm days})}\right)}^{0.38}
\end{equation}

\noindent for Type II SNe, assuming that the absorption $\tau_{\rm 6cm\ peak} \sim1$, from whatever origin, at the time of observed peak in the 6 cm flux density.

The coefficients in Eqs.~14 and 15 depend on the amount of kinetic energy that is transferred to accelerate relativistic electrons and on the details of the acceleration mechanism. Although Fermi acceleration is generally accepted as the relativistic electron acceleration process and it is usually assumed that some fixed fraction of the explosion kinetic energy is transformed into relativistic synchrotron electrons (often assumed $\sim$1\%), the physics of these two aspects is not known in detail {\it a priori}.  Therefore, Eqs.~14 and 15 can only be ``calibrated'' by using the values of well studied RSNe and the assumption that all RSNe of the same type will have similar characteristics.  The constants in Eqs.~14 and 15 have thus been determined from averages within the two RSN subtypes (Type Ib/c and Type II) of those RSNe which have pure, uniform absorption (\ie $K_3 = 0$), with the omission of SN 1987A because of its blue supergiant (BSG) rather than red supergiant (RSG) progenitor.

It should be kept in mind that, because the detailed mechanism of radio emission from SNe is not well understood and the estimates have to rely on {\it ad hoc} calibrations from the few RSNe which have well measured light curves, mass-loss rate estimated from Eqs.~14 and 15 can only complement, and perhaps support, the more accurate determinations done from radio absorption using Eqs.~11 and 13.  

\subsubsection{Changes in Mass-loss Rate}

A particularly interesting case of mass-loss from an RSN is SN1993J, where detailed radio observations are available starting only a few days after explosion.  \cite{VanDyk94} find evidence for a {\it changing} mass-loss rate for the presupernova star which was as high as $\sim10^{-4}\ {\rm M_\odot}\ {\rm yr}^{-1}$ approximately 1000 years before explosion and decreased to $\sim10^{-5}\ {\rm M_\odot}\ {\rm yr}^{-1}$ just before explosion, resulting in a relatively flat density profile of $\rho \propto r^{-1.5}$.

\citet{Fransson98} have suggested that the observed behavior of the f-f absorption for SN1993J could alternatively be explained in terms of a systematic decrease of the electron temperature in the circumstellar material as the SN expands.  It is not clear, however, what the physical process is which determines why such a cooling might occur efficiently in SN1993J, but not in SNe such as SN1979C and SN1980K where no such behavior is required to explain the observed radio turn-on characteristics.  Also, recent X-ray observations with ROSAT of SN 1993J imply a non-$r^{-2}$ CSM density surrounding the SN progenitor \citep{Immler01}, with a density gradient of $\rho \propto r^{-1.6}$ .

Moreover, changes in presupernova mass-loss rates are not unusual.  \citet{Montes00} find that Type II SN1979C had a slow increase in its radio light curve after day $\sim4300$ which implied an {\it excess} in flux density by a factor of $\sim$1.7 with respect to the standard model, or a density enhancement of $\sim30\%$ over the expected density at that radius.  On the other hand, Type II SN1980K showed a steep {\it decline} in flux density at all wavelengths by a factor of $\sim2$ occurring between day $\sim$3700 and day $\sim$4900.  Such a sharp decline in flux density implies a {\it decrease} in $\rho_{\rm CSM}$ by a factor of $\sim1.6$ below that expected for a $r^{-2}$ CSM density profile \citep{Montes98}.  If one assumes the radio emission arises from a $\sim10^4$ \kms blastwave traveling through a CSM established by a $\sim10$ \kms pre-explosion stellar wind, this implies a significant change in the stellar mass-loss rate, for a constant speed wind, $\sim12,000$ yr before explosion for both SNe.

\section{SN1998bw}

The radio emission from SN 1998bw reached an unusually high 6 cm spectral luminosity at peak of $\sim6.7 \times 10^{28}$ erg s$^{-1}$ Hz$^{-1}$, \ie $\sim3$ times higher than either of the well studied, very radio luminous Type IIn SNe SN 1986J and SN 1988Z, and $\sim40$ times higher than the average peak 6 cm spectral luminosity of Type Ib/c SNe.  It also reached this 6 cm peak rather quickly,  only $\sim13$ days after explosion.

SN 1998bw is unusual in its radio emission, but not extreme.  For example, the time from explosion to peak 6 cm luminosity for both SN 1987A and SN 1983N was shorter  and, in spite of the fact that SN1998bw has been called ``the most luminous radio supernova ever observed,'' its  6 cm spectral luminosity  at peak is exceeded by that of SN 1982aa \citep{Yin94}.  However,  SN 1998bw is certainly the most radio luminous Type Ib/c RSN observed so far by a factor of $\sim25$ and it reached this higher radio luminosity very early.

\subsection{Expansion Velocity}

Although unique in neither the speed of radio light curve evolution nor in peak 6 cm radio luminosity, SN1998bw is certainly unusual in the combination of these two factors -- very radio luminous very soon after explosion.  \citet{Kulkarni98a,Kulkarni98b} have used these observed qualities, together with the lack of interstellar scintillation (ISS) at early times, brightness temperature estimates, and physical arguments to conclude that the blastwave from SN1998bw giving rise to the radio emission must have been expanding relativistically.  On the other hand, \citep{Waxman99} argue that a sub-relativistic blastwave can generate the observed radio emission.  However, both sets of authors agree that a very high expansion velocity ($\gtrsim0.3c$) is required for the radio emitting region under a spherical geometry.

Simple arguments confirm this high velocity.  To avoid the well known Compton Catastrophe, \citet{Kellermann69} have shown that the brightness temperature $T_{\rm B} < 10^{12}$ K must hold and \citet{Readhead94} has better defined this limit to $T_{\rm B} < 10^{11.5}$ K.  From geometrical arguments, such a limit requires the radiosphere of SN1998bw to have expanded at an apparent speed $\gtrsim230,000$ \kms, at least during the first few days after explosion.  While such a value is only mildly relativistic ($\Gamma \sim$ 1.6; $\Gamma = \frac{1}{\sqrt{1-\frac{v^2}{c^2}}}$), it is still unusually high.  Measurements by \citet{Gaensler97} and  \citet{Manchester01} have demonstrated that the radio emitting regions of the Type II SN1987A expanded at an {\it average} velocity of $\sim35,000$ \kms over the 3 years from 1987 February to mid-1990 so that, in a very low density environment such as one finds around Type Ib/c SNe, very high blastwave velocities appear to be possible.

\subsection{Radio Light Curves}

An obvious comparison of SN1998bw with other RSNe is the evolution of its radio flux density at multiple frequencies and its description by known RSN models.  The available radio data can be found on \\
{\it http://www.narrabri.atnf.csiro.au/~mwiering/grb/grb980425/} \\  and are plotted in Fig.~\ref{fig2}.  SN1998bw shows an early peak which reaches a maximum as early as day 10 -- 12 at 8.64 GHz, a minimum almost simultaneously for the higher frequencies ($\nu \ge$ 2.5 GHz) at day $\sim$ 20 -- 24, then a secondary, somewhat lower peak at later times after the first dip.  An interesting characteristic of this ``double humped'' structure is that it dies out at lower frequencies and is relatively inconspicuous in the 1.38 GHz radio measurements (see Fig.~\ref{fig2}).

\citet{Li99} propose an initially synchrotron self-absorbed (SSA), rapidly expanding blastwave in a $\rho \propto r^{-2}$ circumstellar wind model to describe the radio light curve for SN 1998bw.  This is in many ways similar to the \citet{Chevalier98} model for Type Ib/c SNe, which also included SSA.  To produce the second peak in the radio light curves \citet{Li99} postulate a boost of blastwave energy by a factor of $\sim2.8$ on day $\sim$22 in the observer's time frame.

Modeling of the radio data for SN 1998bw with the well established formalism for RSNe presented above shows that such an energy boost is not needed.  A fast shock interacting with a dense, slow, stellar wind established, ionized CSM, which is modulated in density over time scales similar to those seen for RSNe, can produce a superior fit to the data.  No blastwave reacceleration is required and no synchrotron self-absorption (SSA) at early times is apparent.  The parameters of our best fit model are given in Table  \ref{tbl-1} and shown as the curves in Fig.~\ref{fig2}.  \citet{Li99} do not give a quantitative indication of the quality of their fit.  However, a visual comparison of the curves in Fig.~\ref{fig2} with those of \citet{Li99} Fig. 9, shows that our purely thermal absorption model with structured CSM provides a superior fit.

\section{Discussion}

\subsection{Radio Light Curve Description}

An unusual feature of the SN1998bw radio light curves shown in Fig.~\ref{fig2} is its apparently ``double humped'' structure with a second peak in the radio emission near Day $\sim30$.  Such a ``double humped'' structure of the radio light curves can be reproduced by a single energy blastwave encountering differing CSM density regimes as it travels rapidly outward.  This is a reasonable assumption since complex density structure in the CSM surrounding SNe, giving rise to structure in the radio light curves, is very well known in such objects as SN1979C \citep{Weiler91,Weiler92a,Montes00}, SN1980K \citep{Weiler92b,Montes98}, SN1988Z \citep{VanDyk93,Lacey01} and, particularly, SN1987A \citep{Jakobsen91}.

Additionally, what has not been previously recognized is a sharp drop in the radio emission near Day $\sim75$ and a single measurement epoch at Day 192 which is significantly ($\sim60\%$) higher at all frequencies than expected from the preceeding data on Day 149 and the following data on Day 249.

Quite surprisingly, both of these temporary increases in radio emission can be explained by the SN blastwave encountering physically similar shells of enhanced density.  The first enhancement or ``bump'' after the initial outburst peak is estimated to start on Day 25 and end on Day 75, \ie having a duration of $\sim50$ days and turn-on and turn-off times of about 12 days, where the increased radio emission and absorption ($K_1$ and $K_3$) by a factor of $\sim1.6$ imply a density enhancement of $\sim30\%$.  Exactly the same density enhancement factor and length of enhancement is compatible the ``bump'' observed in the radio emission at Day 192 (\ie the single measurement within the 100 day gap between measurements on Day 149 and Day 249), even though the logarithmic time scale of Fig.~\ref{fig2} makes the time interval look much shorter.  Unfortunately, the decreased sampling interval has only one set of measurements altered by the proposed Day 192 enhancement, so that we cannot determine its length other than being $<100$ days.

Note that the fit listed in Table \ref{tbl-1} and shown as the curves in Fig.~\ref{fig2} requires no ``uniform'' absorption ($K_2 = 0$) so that all of the f-f absorption is due to a clumpy medium as described in Eqs.~1 and 5.  These results, combined with the estimate of a high blastwave velocity, suggest that the CSM around SN1998bw is highly structured with little, if any, inter-clump gas.  The clump filling factor has to be high enough to intercept a considerable fraction of the blastwave energy and low enough to let radiation escape from any given clump without being appreciably absorbed by any other clump, i.e., a filling factor of $\sim$0.1 -- 0.3. The blastwave can then easily move at a speed which is a significant fraction of the speed of light, because it is moving in a very low density medium, but still cause strong energy dissipation and relativistic electron acceleration at the clump surfaces facing the SN explosion center. 

 If the knots are dense and relatively opaque to radio emission, we mainly observe radiation produced at the surfaces of the knots in the CSM on the far-side of the SN in which some internal absorption may also be occurring. The radiation from near-side knots is probably absorbed internally by the knots themselves and, therefore, lost from the signal we detect. 

Note also from the fit given in Table \ref{tbl-1} that the presence of a $K_4$ factor implies there is a more distant, uniform screen of ionized gas surrounding the exploding system which is too far to be affected by the rapidly expanding blastwave and provides a time independent absorption.

\subsection{Physical Parameter Estimates}

Using the fitting parameters from Table \ref{tbl-1} and Eq. 13, we can estimate a mass-loss rate from the pre-explosion star.  However, the proper parameter assumptions are rather uncertain for these enigmatic objects.  For a preliminary estimate we shall assume $t_{\rm i} = 23~{\rm days}$, $t = (t_{\rm 6cm\ peak} - t_0) = 13.3$ days, $m = -(\alpha - \beta - 3)/3 = 0.78$ (Eq.~7), $w_{\rm wind}$ = 10 \kms (for an assumed RSG progenitor), $v_{\rm i} = v_{\rm blastwave} = 230,000$ \kms, and $T = 20,000$ K.  We shall also assume, since the radio emission implies that the CSM is highly clumped (\ie $K_2 = 0$), a filling factor, $f$, of only $f \sim10$\%, which reduces the estimated mass loss rate by a factor of $\sqrt{f}$.  Within these rather uncertain assumptions, Eq.~13 yields an estimated mass-loss rate of ${\dot M} \sim3.5 \times 10^{-5}\ {\rm M_\odot}\ {\rm yr}^{-1}$ with density enhancements of $\sim\sqrt{1.6}$ or $\sim30\%$  during the two known, extended ``bump'' periods.

Assuming that the blastwave is traveling at a constant speed of $\sim230,000$ \kms  the start of the first ``bump'' on Day 25 implies that it starts at $\sim5.0 \times 10^{16}$ cm and ends on Day 75 at $\sim1.5 \times 10^{17}$ cm from the star.  Correspondingly, if it was established by a 10 \kms RSG wind, the 50 days of enhanced mass-loss ended $\sim1,600$\ yr and started $\sim4,700$\ yr before the explosion.  The earlier high mass-loss rate epoch indicated by the enhanced emission on Day 192 in the measurement gap between Day 149 and Day 249 implies, with the same assumptions, that it occurred in the interval between $\sim9,400$ yr and $\sim15,700$ yr before explosion.  It is interesting to note that the time between the presumed centers of the first and second increased mass-loss episodes of $\sim9,400$\ yr is comparable to the $\sim12,000$\ yr before explosion at which SN1979C had a significant mass-loss rate increase \citep{Montes00} and SN1980K had a significant mass-loss rate decrease \citep{Montes98}, thus establishing a possible characteristic time scale of $\sim10^4$\ yr for significant changes in mass-loss rate for pre-explosion massive stars.  As discussed by \citet{Panagia01} and \citet{Bono01}, mass-loss variations with such time scales can be explained in terms of pulsational instabilities of red supergiants with initial masses of $\sim20\ {\rm M_\odot}$.

\section{GRB Radio Afterglows and Type Ib/c RSNe}

Since the suggestion of a possible relation between SN1998bw and GRB980425, it has remained a tantalizing possibility that the origin of at least some GRBs is in the better known Type Ib/c SN phenomenon.  First, of course, one must keep in mind that there may be (and probably is) more than one origin for GRBs, a situation which is true for most other classes of objects.  For example, SNe, after having been identified as a new phenomenon in the early part of the last century, quickly split into several subgroups such as Zwicky's Types I -- VI, then coalesced back into just two subgroupings based on H$\alpha$ absent (Type I) or H$\alpha$ present (Type II) in their optical spectra.  This simplification has not withstood the test of time, however, and subgroupings of Type Ia, Ib, Ic, II, IIb, IIpec, IIn, and others have come into use over the past 20 years.

GRBs, although at a much earlier stage of understanding, have similarly started to split into subgroupings.  The two currently accepted groupings are referred to as ``fast-hard'' and ``slow-soft'' from the tendency of the $\gamma$-ray emission for some to evolve more rapidly (mean duration $\sim$0.2 s) and to have a somewhat harder spectrum than for others which evolve more slowly  (mean duration $\sim$20 s) with a somewhat softer spectrum \citep{Fishman95}.

Since we are only concerned with the radio afterglows of GRBs here, all of our examples fall into the ``slow-soft'' classification, at least partly because the ``fast-hard'' GRBs fade too quickly for followup observations to obtain the precise positional information needed for identification at longer wavelengths.  It is therefore uncertain at present whether ``fast-hard'' GRBs have radio afterglows or even whether the ``slow-soft'' GRBs represent a single phenomenon.  If, however, we assume that all ``slow-soft'' GRBs have a similar origin and that GRB980425/SN1998bw is a key to this puzzle telling us that  ``slow-soft'' GRBs have their origin in Type Ib/c SNe, we can investigate relations between the two observational phenomena.

Keeping in mind that much of the radio data available on Type Ib/c SNe is of poor quality due to their relative faintness, and that the situation is even worse for GRB radio afterglows, some of the similarities are:

\begin{enumerate}
\item{GRB980425 and Type Ib SN1998bw are possibly related.}
\item{Both GRBs and Type Ib/c SNe are suggested by some models to originate from the core collapse of a massive (perhaps stripped) progenitor star.}
\item{Both GRBs and Type Ib/c SNe are suggested, by some models, to release $\sim10^{51} - 10^{52}$ ergs of kinetic energy in a few seconds.}
\item{Both GRBs and Type Ib/c SNe are detectable at cm wavelengths very soon after the explosion -- within a few days, while Type II SNe generally have much longer delays.}
\item{Both GRBs and Type Ib/c RSNe generally have relatively steep radio spectra  ($\alpha \lesssim -1$; $S \propto \nu^{+\alpha}$) compared to, for example, extragalactic QSOs or radio galaxy nuclei ($\alpha \gtrsim -0.75$) or Type II RSNe ($\alpha \gtrsim -1$).}
\end{enumerate}

There is presently one clear difference between known GRBs and Type Ib/c SNe, however.  Type Ib/c RSNe can generally be treated as classical, non-relativistic phenomena, sufficiently close that cosmological effects can be neglected, while most known GRBs appear to be highly relativistic phenomena ($\Gamma \sim$ 10) at cosmological distances (z $\sim$ 1).  SN1998bw/GRB980425, which is not very distant (z = 0.0085, $\sim$40 Mpc) and only mildly relativistic ($\Gamma \sim1.6$), is possibly the ``bridge'' between the two groups.  We shall discuss the properties of specific GRBs in more detail in a following paper.

While observational similarity does not necessarily mean identical physical processes, the possibility of similar or related phenomena giving rise to both GRBs and Type Ib/c SNe must be considered.  Speculatively, GRBs may represent a rare subclass of Type Ib/c SNe and/or the subset of Type Ib/c SNe where we lie within the narrow emission cone of a relativistic jet.  In either case, they are sufficiently rare that they can only be found by sampling very large volumes of space, i.e., preferentially at high redshifts.

\section{Summary} 

We have collected and rationalized a parameterization and physical model description which has been very successful in explaining the radio emission from SNe.  This we have then applied to SN1998bw, which is thought to be related to GRB980425.  From this parameterization of the radio emission for SN1998bw, we are able to describe the radio light curves for SN1998bw in previously unattainable detail and show that a mildly relativistic blastwave interacting with a clumpy, structured CSM established by the wind from a massive progenitor provides a satisfactory description.  Superimposed on a relatively constant mass-loss of $\sim3.5 \times 10^{-5}\ {\rm M_\odot}\ {\rm yr}^{-1}$ are epochs of higher mass-loss rate which increase the CSM density by a factor of $\sim30\%$.  At least two of these episodes can be identified and occur at an interval of $\sim9,400$\ yr and a duration of $\sim3,000$\ yr.

\acknowledgements

KWW \& MJM wish to thank the Office of Naval Research (ONR) for the 6.1 funding supporting this research.  Additional information and data on radio emission from SNe and GRBs can be found on {\it http://rsd-www.nrl.navy.mil/7214/weiler/} and linked pages.

\clearpage

\begin{deluxetable}{cc}
\tablewidth{0pt}
\tablecaption{SN1998bw Modeling Results. \label{tbl-1}}
\tablehead{
\colhead{Parameter}         & \colhead{Value} }
\startdata
$\alpha$            & -0.71 \\ 
$\beta$             & -1.38 \\ 
$K_1$\tablenotemark{a}               & $2.4 \times 10^3$ \\  
$K_2$               & 0  \\
$\delta$            & -- \\
$K_3$\tablenotemark{a}               & $1.7 \times 10^3$ \\
${\delta}^{\prime}$ & -2.80 \\
$K_4$               & $1.2 \times 10^{-2}$ \\
$t_0$(Explosion Date) & 1998 Apr. 25.90915 \\
($t_{\rm 6cm\ peak} - t_0$)(days) & 13.3 \\
$S_{\rm 6cm\ peak}$(mJy) & 37.4 \\
d(Mpc) & 38.9 \\
$L_{\rm 6cm~peak}\ ({\rm ergs\ s^{-1}\ Hz^{-1}})$ & $6.7 \times 10^{28}$ \\
${\dot M}({\rm M_\odot} ~ {\rm yr}^{-1})$\tablenotemark{b} & $3.5 \times 10^{-5}$ \\
\enddata
\tablenotetext{a}{Enhanced by a factor of $1.6$ over the intervals Day 25 - 75 and Day 165 - 215, although the latter interval could be as long as 100 days and still be compatible with the available data.}
\tablenotetext{b}{Assuming $t_{\rm i} = 23~{\rm days}$, $t = (t_{\rm 6cm\ peak} - t_0) = 13.3$ days, $m = -(\alpha - \beta - 3)/3 = 0.78$, $w_{\rm wind}$ = 10 \kms, $v_{\rm i} = v_{\rm blastwave} = 230,000$ \kms, $T = 20,000$ K, and filling factor, $f = 0.1$.}
\end{deluxetable}

\clearpage

\figcaption[newfig_v2.eps]{Cartoon, not to scale, of the SN and its shocks along with the stellar wind established CSM and more distant ionized material. The radio emission is thought to arise near the blastwave.  The expected locations of the several absorbing terms in Eqs.~1 -- 10 are illustrated.\label{fig1}}

\figcaption[fig98bw4bw_5Apr2001.ps]{The radio light curves of SN 1998bw at 8.64 GHz (3.5 cm; upper left, {\it open circles}, {\it solid line}), 4.80 GHz (6.3 cm; upper right, {\it stars}, {\it dashed line}), 2.50 GHz (12 cm; lower left, {\it open squares}, {\it dash-dot line}) and 1.38 GHz (21.7 cm; lower right, {\it open triangles}, {\it dotted line}).  The curves are derived from a best fit model described by the Eqs. 1-10 and the parameters and assumptions listed in Table \ref{tbl-1}. During the 50 day intervals from Day 25 - 75 and from Day 165 - 215 the emission and absorption factors ($K_1$ and $K_3$) are increased by a factor of 1.6 with a 6 day boxcar smoothed turn-on and turn-off of the enhanced emission/absorption. The physical interpretation of these enhancements is discussed in $\S$4. \label{fig2}}

\clearpage

\begin{figure}
\figurenum{fig1}
\epsscale{0.9}
\plotone{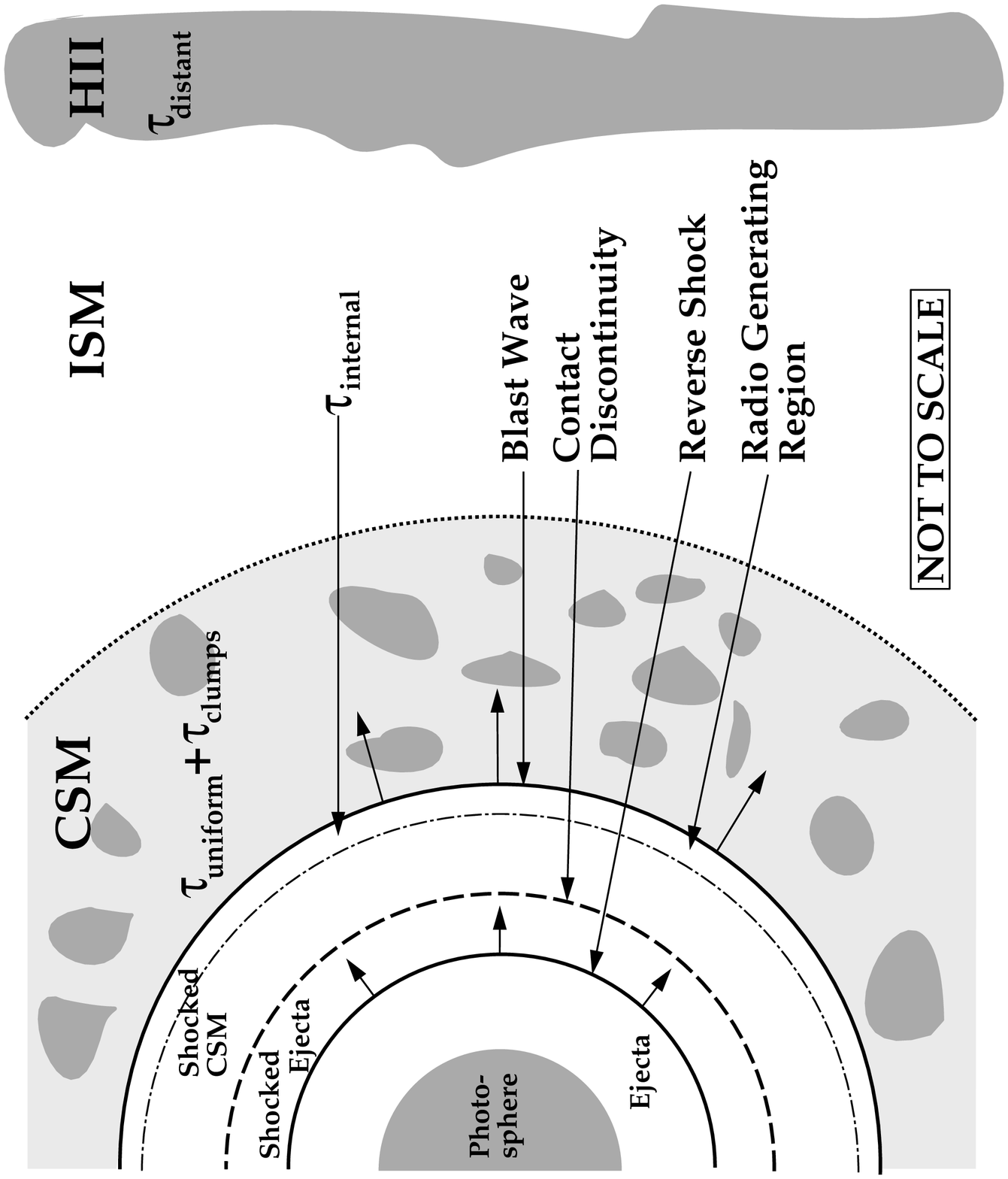}
\end{figure} 

\clearpage

\begin{figure}
\figurenum{fig2}
\epsscale{1.0}
\plotone{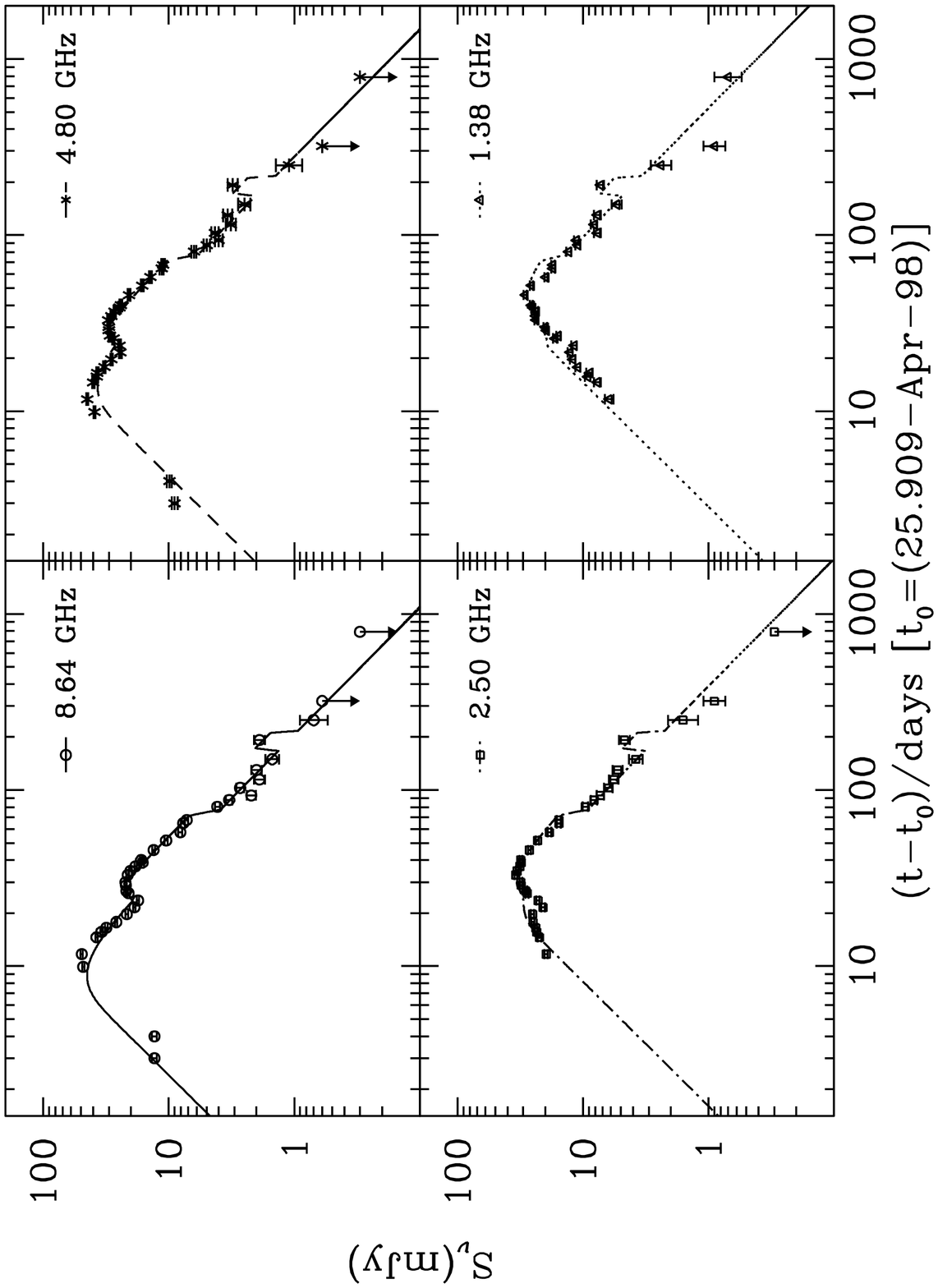}
\end{figure}

\end{document}